# Using Pi-calculus to Model Dynamic Web Services Composition Based on the Authority Model

Sok-Min Han, ,
College of Computer Science, **KIM IL SUNG** University, Pyongyang, D.P.R.K
Un-Chol Pang
College of Computer Science, **KIM IL SUNG** University, Pyongyang, D.P.R.K
Hyok-Chol Choe
College of Computer Science, **KIM IL SUNG** University, Pyongyang, D.P.R.K
Chol-Jun Hwang
Electronic Library. **KIM IL SUNG** University, Pyongyang, D.P.R.K

*Abstract* – There are lots of research works on web service, composition, modeling, verification and other problems. Theses research works are done on the basis of formal methods, such as petri-net, pi-calculus, automata theory, and so on. Pi-calculus is a natural vehicle to model mobility aspect in dynamic web services composition (DWSC). However, it has recently been shown that pi-calculus needs to be extended suitably to specify and verify DWSC.

In this paper, we considers the authority model for DWSC, extends pi-calculus in order to model dynamic attributes of system, and proposes a automatic method for modeling DWSC based on extended pi-calculus.

## 1. Introduction

There are lots of research works on web service, composition, modeling, verification and other problems. Theses research works are done on the basis of formal methods, such as petri-net, pi-calculus, automata theory, and so on.

Web Services are web applications which are obtained through the web to satisfy the needs of the user [1]. These applications are autonomous computer programs. Using a single web service alone cannot satisfy all the requirements of the user. Hence, there is a need to compose multiple web services to satisfy the complex needs. This process of combining together various web services is called Web Service Composition [1,2]. Web Service Compositions are classified broadly into two types depending on the time of service selection for composition. If the service selections are done at design time or compile time then this is called Static Web Service Composition. In the DWSC, the services are selected at run time. The main advantage of DWSC technique is that web services can be dynamically discovered and invoked on demand [2].

The best important problems related to DWSC are to ensure their correctness, since the number of web services is increasing rapidly.

Therefore we have to model formally dynamic web service composition and verify using model checking tool.

Current verification approaches are based on Process Algebra, Petri-net and Automata theory[3,4,5].

In [6], to describe the web service composition, Finite State Automata (FSM) is used. Then using Promela language, the FSM is translated into programs and finally verified using model checking tool SPIN [7,8]. This technique cannot be used in the presence of mobility. In [10,11,12,14], to describe the web service composition, Calculus of Communicating System (CCS) is used. CSS cannot handle the mobility that is transferring channels or links. For this reason, Pi-Calculus is started to use to overcome that problem.

We have used pi-calculus for describing, modeling and verifying DWSC, pi-calculus needs to be extended suitably to specify and verify DWSC.

The rest of the paper is organized as follows. Section 2 proposes framework of DWSC based on the authority model, and section 3 extends pi-calculus for modeling proposed framework. Extended pi-calculus based modeling and conclusions are given in section 4 and 5, respectively.

## 2. Related work

In [13], it mainly introduces how to use Pi-calculus to describe web services, and do the corresponding validation and analysis. It expresses web service semantics and further describes the web service composition in view point of the process. Pi-calculus is regarded as the web service modeling language. This approach mainly considers the Pi-calculus description of web services, but it lacks of the research about service interaction and composition.

In [9,18], it proposes a method of describing and verifying web services using Pi-calculus. This method discusses some difference among the Pi-calculus and other formal methods, and it illustrates relationship between Pi-calculus and web services protocol stack. The rules about constructing web service composition model using Pi-calculus and how to find agent and channel are all described. The service composition discussed in this approach involves the interaction between service and user, service and service, but it does not consider the corresponding user's role and the request goal about service collaboration.

In [21], it has shown how Pi-calculus can be effectively be used for HMS (a hospital management system). It has designed the architecture of the HMS and modeled the relevant web services, and obtained the composed web service in the advent of failures. Finally it has verified the composed services using the tool MWB.

In [22], it mainly describes the process of modeling service interaction modes using Pi-calculus through a practical case of express parcels. This web service composition involves 5 services: service of express delivery company, express delivery query service, map service, RFID integrity testing service and weather service. Then the service has expressed as process node in Pi-calculus, and the interaction information between services has expressed as process channel.

In [23], it proposes a technique for dynamic web service composition that is value based and provides composed web services based on QoS. Value meta-model and its representation language VSDL(Value-based Service Description Language) are presented. Values are used to define quality of web services. Value added service broker architecture is proposed to dynamically compose the web services and value-meta model to define relationship among values.

In [24], it proposes a framework for automated web services composition through AI(Artificial Intelligence) planning technique by combining logical combination and physical composition. This paper discusses the real life problems on the web that is related to planning and scheduling.

Hence these problems doesn't relate to formal verification.

## 3. Framework of Dynamic Web Service Composition

The dynamic web service composition methods generate the request/response automatically. First request goes to Service Discovery, and then the services are selected from web service database (WSDB) that meet user criteria. Now the Composition Generation composes these services. If there is more than one composite service that meets user criteria then Composition Selection evaluates them and returns the best selected service to Composition Execution. The results are returned to requester.

Framework of DWSC is shown in Figure 1.

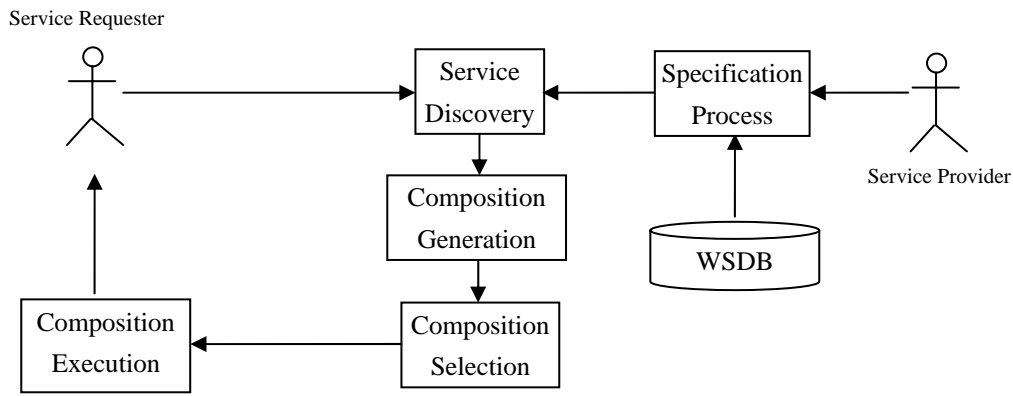

Figure 1.Framework of DWSC

Framework of DWSC consists of following modules as shown in Figure 1.
- Specification Process

It is the process of specification of web services to the system. New services are registered in WSDB through Specification Process.
- Service Discovery

If requester send request, then result in comparing service specification *spec* and request specification *spec'*. Semantic technique is used to dynamically discover web service.
- Composition Generation

The purpose of Composition Generation is the process to generate web service composition for satisfying the request goal based on interaction between discovered web services.
- Composition Selection

Composition Selection evaluates selected web services and returns the best selected service based on specified criteria.
- Composition Execution

Composition Execution executes above web services and results are sent back to requester.

## 4. Syntax and Semantics of Pi-calculus

A process algebra called pi-calculus can be used to show the interaction with the system. A process algebra is a formalism used to specify the behavior of systems in a modular and precise way. There are two fundamental units associated with pi-calculus namely process and name. Concurrent entities are dented by process and names are transferred during the communication between processes. Atomic services communicate with each other by sending and receiving messages to form a composite service. Pi-calculus has string reasoning ability. To verify the correctness of a composite system, a lot of related tools are being developed by the researchers[14].

A process *S* can be defined as follows;

$$S ::= 0 \mid P \cdot Q \mid c'\langle x \rangle.P \mid c(x).P \mid \tau.P \mid P+Q \mid P||Q \mid (vc)P \mid [x=y]P$$

(1) *0* (null process) is inactive process, which doesn't execute any operation and can also, be expressed as NIL.

(2) *P · Q* (sequence composition) represents the sequential execution of processes *P* and *Q*. This means that first *P* will execute followed by the execution of *Q*.

(3) *c'⟨x⟩.P* (input process) denotes to send *x* into channel *c* and then execute process *P*.

(4) *c(x).P* (output process) denotes to receive *x* from channel *c* and then execute process *P*.

(5) *τ.P* (silent process): denotes to execute process P directly. *τ* is a dummy action, which do nothing.

(6) *P+Q* (non-deterministic choice composition) represents non-deterministic selection of either *P* or *Q* for the execution.

(7) *P||Q* (parallel composition) represents that *P* and *Q* are concurrently executed. *P* and *Q* can exchange messages through channel.

(8) *(vc)P* (restriction) denotes that *P* cannot communicate with environment through channel *c*, but communication through path *c* can go on inside *P*.

(9) *[x=y]P* (conditions) denotes that if name *x* is equal to *y*, then execute process *P*.

The terms (1)-(9) is used for making of verification model.

Considering the syntax of Pi-calculus, we may consider its operational semantics.

Based on the operational semantics of Pi-calculus, we precisely may understand the action performed by process; the operational semantics of Pi-calculus is the set of conditional rules denotes processes may perform an action. The conditional rule is of form $\frac{A}{B}$. This denotes B is true if A is true. Expression $P \xrightarrow{\alpha} Q$ denotes $P$ can perform $\alpha$ and become $Q$.

Therefore, operational semantics of Pi-calculus can be defined as follows;

(1) $$\frac{}{\alpha.P \xrightarrow{\alpha} P}$$

The $\alpha.P$ performs the action $\alpha$ and then behaves as process $P$.

(2) $$\frac{P \xrightarrow{\alpha} P'}{P+Q \xrightarrow{\alpha} P'}$$

If $P$ can perform $\alpha$ and become $P^{'}$, then $P + Q$ can perform $\alpha$ and become $P^{'}$.

(3) $$\frac{P \xrightarrow{\alpha} P'}{[x=x]P \xrightarrow{\alpha} P'}$$

If $P$ can perform $\alpha$ and become $P'$, then $[x=x]P$ can perform $\alpha$ and become $P'$.

(4) $\dfrac{P \xrightarrow{\alpha} P'}{P|Q \xrightarrow{\alpha} P'|Q}$, $(bn(\alpha) \cap fn(Q) = \phi)$

If a bound name is run parallel to a free name, this rule holds. If $P$ can perform $\alpha$ and become $P'$, then $P|Q$ can perform $\alpha$ and become $P'|Q$.

(5) $\dfrac{P \equiv P' \wedge P \xrightarrow{\alpha} Q \wedge Q \equiv Q'}{P' \xrightarrow{\alpha} Q'}$

This rule denotes that the semantics of processes identified is equal.

(6) $\dfrac{P \xrightarrow{\alpha} P' \wedge c \notin fn(\alpha) \cup bn(\alpha)}{(vc)P \xrightarrow{\alpha} (vc)P'}$

If $P$ performs $\alpha$ and becomes $P'$, and the name set of action $\alpha$ involves a bound name $c$, then $(vc)P$ performs $\alpha$ and becomes $(vc)P'$. And the bound name $c$ of the process $P$ restricts in process $P'$.

## 5. Framework of DWSC based on the authority model and Extended Pi-calculus

### 5.1. The authority model

According to the roles of service requester, the authorities determines accessed to the web service system. Each authority constraints the roles of service requester and each roles constraints the function of system.

**(Definition 1)** Service requester based authority model is followed as:

$PM:=(UA, P, PA, Op, Ob)$ (1)

Where $UA: U \rightarrow R$,

$P \subseteq 2^{OPS \times OBJ}$

$PA: R \rightarrow P$,

$Op: P \rightarrow 2^{OPS}$

$Ob: P \rightarrow 2^{OBJ}$

$U=\{u_1, u_2, ..., u_n\}$ represents a set of services requesters,

$R=\{r_1, r_2, ..., r_m\}$ represents a set of requester's roles,

$OBJ=\{obj_1, obj_2, ..., obj_k\}$ represents a set of system objects.

$OPS=\{op_1, op_2, ..., op_p\}$ represents a set of operations along to objects.

According to authority $u$ of service requester, the role and goal set is determined as following:

- Service requester's role set

$R_u=UA(u)=\{r_1, r_2, ..., r_n\}$, $R_u \subseteq R$

- Goal set

$Ob_u = Ob(PA(UA(u))) = \{ob_1, ob_2, ..., ob_p\}$, $Ob_u \subseteq OBJ$

**(Definition 2)** The authority model based web service is defined as:

$S := \{(role, goal) | role \in R_u, goal \in Ob_u\}$

## 5.2. Framework of DWSC based on the authority model

The dynamic web service composition methods generate the request/response automatically. First request goes to Authority Process, it determine requester' roles and goals from an authority Information database (AIDB), and then the services are selected from web service database (WSDB) that meet user criteria. Now the Composition Generation composes these services. If there is more than one composite service that meets user criteria then Composition Selection evaluates them and returns the best selected service to Composition Execution. The results are returned to requester.

Framework of DWSC based on the authority model is shown in Figure 1.

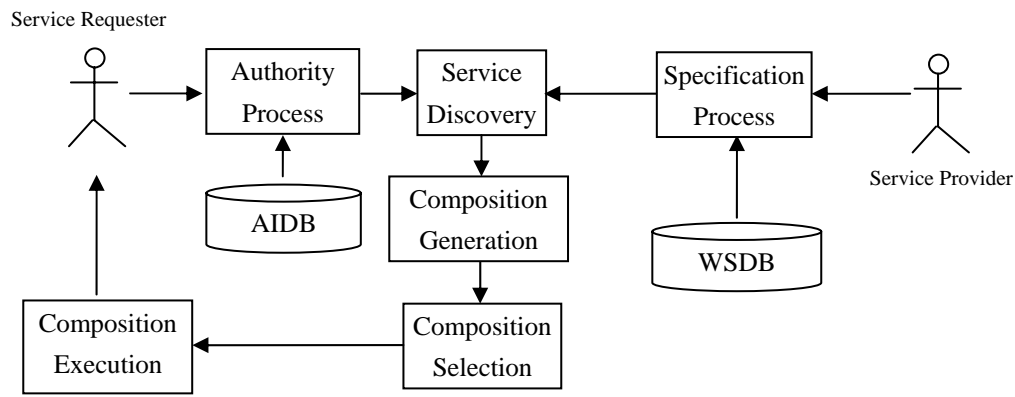

Figure 2. Framework of DWSC based on the authority model

Framework of DWSC based on the authority model consists of following modules as shown in Figure 1.

- Specification Process

It is the process of specification of web services to the system. New services are registered in WSDB through Specification Process.

- Service Discovery

If requester send request, then result in comparing service specification *spec* and request specification *spec'*. Semantic technique is used to dynamically discover web service.

- Authority Process

The purpose of authority Process is to determine service requester's role and goal set with an authority.

- Composition Generation

The purpose of Composition Generation is the process to generate web service composition for satisfying the request goal based on interaction between discovered web services.

- Composition Selection

Composition Selection evaluates selected web services and returns the best selected service based on specified criteria.

- Composition Execution

Composition Execution executes above web services and results are sent back to requester.

### 5.3. Extended Pi-Calculus

We have used pi-calculus for describing, modeling and verifying DWSC, pi-calculus needs to be extended suitably to specify and verify DWSC.

(**Definition 3**) The syntax of extended Pi-calculus can be defined as follows by BNF.

$$S ::= signal(t).P \mid [P,Q]_t$$

where $t=1,2,3,\cdots$

(1) *signal(t).P* is a signal process, which means to produce a signal $t$ that can be received by the process $P$.
(2) $[P,Q]_t$ denotes an event process structure, which means to execute $P$ under normal conditions and switch to execute $Q$ when receiving the signal $t$.

(**Definition 4**) Web services composition on service $S_1$ and $S_2$ is modeled in extended pi-calculus as following:

$$WSC = S_1 \cdot S_2 \mid c'\langle x\rangle.S \mid c(x).S \mid S_1+S_2 \mid S_1||S_2 \mid signal(t).S \mid [S_1,S_2]_t \mid WSC$$

where $S_1 \cdot S_2$ represents the sequential execution of services $S_1$ and $S_2$, $S_1||S_2$ represents that $S_1$ and $S_2$ are concurrently executed.

$S_1+S_2$ represents non-deterministic selection of either $S_1$ or $S_2$ for the execution.

*signal(t).S* denotes to produce a signal $t$ that can be received by the service $S$.

$c'\langle x\rangle.S$ denotes to send $x$ into channel $c$ and then execute service $S$.

$c(x).S$ denotes to receive $x$ from $c$ and then execute service $S$.

$[S_1,S_2]_t$ denotes to execute $S_1$ under normal conditions and switch to execute $S_2$ when receiving the signal $t$.

### 6. The modeling of proposed framework in extended pi-calculus

### 6.1. The modeling process

In order to model the proposed framework using extended pi-calculus is done as following.
(1) Web service is a process in extended Pi-calculus.

Each process of the proposed framework considers web services with roles and goals.
Service set is determined as following;

$S=S_{main} \cup S_{base,}$

$S_{main}=\{S_{registration}, S_{discovery}, S_{generation}, S_{selection}, S_{execution}, S_{authority}\}$,
$S_{base}=\{S_1,S_2,...,S_m\}$,

where $S_{main}$ is a set of services in proposed framework, $S_{base}$ is a set of services made by services provider and $S_k=(role, goal)$, $S_k \in S$, $S_{registration}$ is a service for registering specification of web services in the WSDB, $S_{discovery}$ is one for finding desired services, $S_{generation}$ is one for composing service, $S_{selection}$ is one for selecting one of some services, $S_{execution}$ is one for executing service composition, $S_{authority}$ is one for obtaining user's authority information.

(2) Web service interaction is a channel in Pi-calculus.

$Ch=\{(S_j, ChName_c, S_j, Inf_{ij}^c, Inf_{ji}^c) \mid S_i, S_j \in S, c=1,2,...,n\}$

where $Inf_{ij}^c$ denotes the information transformed from $S_i$ to $S_j$, and $Inf_{ji}^c$ denotes the information transformed from $S_j$ to $S_i$.

To model Web service interaction based on Pi-calculus is followed picture.

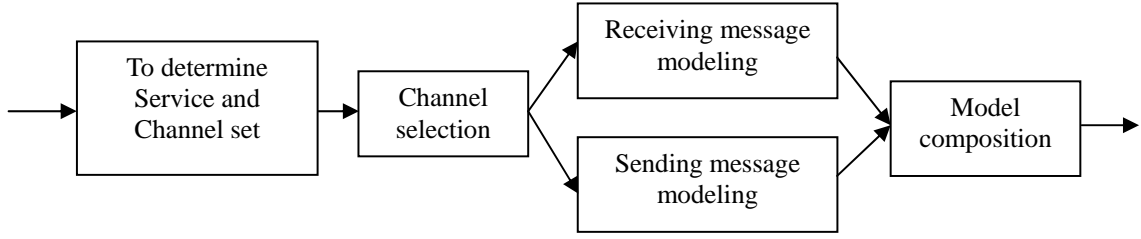

Figure3. The modeling process

1) To determine service and channel set.

Due to every service in Web service composition, its role and goal has considered and its interaction has determined.

$S=\{S_1,S_2,...,S_m\}$, $Ch=\{Ch_1,Ch_2,...,Ch_n\}$, $U=\{u_1,u_2,...,u_n\}$

*About each $u_i \in U$,*

$S_k=(role,goal)$, $role \in R_u$, $goal \in Ob_u$,

$Ch_c=(S_i, ChName_c, S_j, Inf_{ij}^c, Inf_{ji}^c)$

2) Channel selection

On every channels whether is sending message or receiving message along with direction of the message transformation.

3) Receiving message modeling

$Ch_c'\langle S_k.role \rangle.Ch_c(msg)$

*If $msg=S_k.role$ then $.[msg=S_k.role](Ch_c'\langle S_k.goal \rangle.Ch_c(msg))$*

*If $msg=S_k.goal$ then $.[msg=S_k.goal](Ch_c'\langle Inf_{ij}^c \rangle.Ch_c(Inf_{ji}^c))$*

4) Sending message modeling

$Ch_c(msg)$

*If $msg=S_k.role$ then $.[msg=S_k.role](Ch_c'\langle S_k.role \rangle.Ch_c(msg))$*

*If $msg=S_k.goal$ then $.[msg=S_k.goal](Ch_c'\langle S_k.goal \rangle.Ch_c(msg))$*

*If $msg=Inf_{ij}^c$ then $.[msg=Inf_{ij}^c]Ch_c'\langle Inf_{ji}^c \rangle$*

5) Model composition

It has made of a model by composing modeling result of receiving message and sending message.

**6.2. The modeling of DWSC based on extended pi-calculus**

In this section, we model DWSC using extended Pi-calculus. DWSC involves the following services: $S_{registration}$, $S_{discovery}$, $S_{generation}$, $S_{selection}$, $S_{execution}$, $S_{authority}$, $S_{requester}$, $S_{provider}$. $S_{requester}$ and $S_{provider}$ consider a web service for request and providing, respectively. The logic execution order among these services are shown in figure 3.

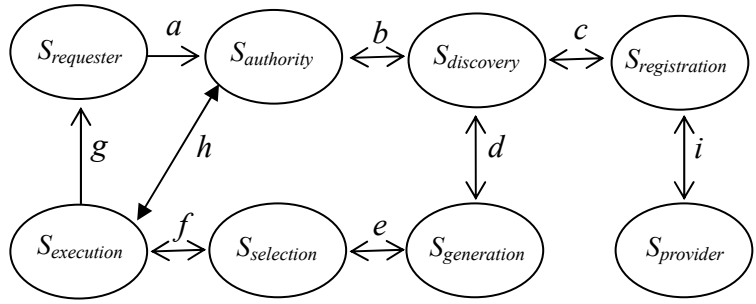

Figure 4. The diagram of interaction between web services

In the above figure 4, *a* is the channel between $S_{requester}$ and $S_{authority}$. *b* is the channel between $S_{discovery}$ and $S_{authority}$. *c* is the channel between $S_{discovery}$ and $S_{registration}$. *d* is the channel between $S_{discovery}$ and $S_{generation}$. *e* is the channel between $S_{generation}$ and $S_{selection}$. *f* is the channel between $S_{selection}$ and $S_{execution}$. *g* is the channel between $S_{requester}$ and $S_{execution}$. *i* is the channel between $S_{provider}$ and $S_{registration}$.

According to the above process, extended Pi-calculus based on DWSC modeled as following.

$DWSC=S_1|S_2$

$S_1=S_{provider}\cdot S_{registration}$, $S_{provider}=\tau.S_{provider}+a'\langle S_{provider}.role\rangle.S_{provider}$

$S_{registration}=a(msg).[msg=S_{provider}.role]S_{registration}+b'\langle S_{registration}.goal\rangle.S_{registration}+signal(t).S_{registration}$

$S_2=S_{requester}\cdot S_{discovery}\cdot S_{generation}\cdot S_{selection}\cdot S_{execution}\cdot S_{authority}$

$S_{requester}=\tau.S_{requester}+c'\langle S_{requester}.role\rangle.S_{requester}$

$S_{discovery}=c(msg).[msg=S_{requester}.role]S_{discovery}+d'\langle S_{discovery}.goal\rangle.S_{discovery}+b(msg1).[msg1=S_{registration}.goal]S_{discovery}+b'\langle S_{requester}.goal\rangle.S_{discovery}$

$S_{generation}=d(msg).[msg=S_{discovery}.goal]S_{generation}+e'\langle S_{generation}.goal\rangle.S_{generation}+[S_{generation}, S_{discovery}]_t$

$S_{selection}=e(msg).[msg=S_{generation}.goal]S_{selection}+f'\langle S_{selection}.goal\rangle.S_{selection}+[S_{selection}, S_{discovery}]_t$

$S_{execution}=f(msg).[msg=S_{selection}.goal]S_{execution}+[S_{execution}, S_{discovery}]_t$

## 7. Conclusion

In the modern service-oriented software engineering, Web services and the formal modeling and verification of DWSC are important research issues.

In this paper, we considered the authority model for DWSC, extended pi-calculus in order to model dynamic attributes of system, and proposed a method for modeling DWSC based on extended pi-calculus.

The result of modeling are used in verified model, we can verify correctness of the system using this models.

Interaction, Journal of Computational Information Systems, 2013, pp.1759-1767.

[4] L.Kuang, Y.J.Xia, S.G.Deng and J.Wu, Analyzing behavioral substitution of Web services based on Pi-calculus, IEEE International Conference on Web Services, 2010, pp.441-448.

[5] P.Petros and F.Jacques, Formal verification of Web services composition using Linear Logic and the Pi-calculus, IEEE European Conference on Web Services, 2011, pp.31-38.

[6] K.Q.He, J.Wang and P.Liang, Semantic interoperability aggregation in aervice requirements refinement, Journal of Computer Science and Technology, vol.25, 2010, pp.1103-1117.

[7] J.X.Liu, K,Q,He, and D.Ning, A Kind of On-demand Web Services Selection Method Based on RGPS, Journal of Computational Information Systems, 2011, 7(1), pp.1-8.

[8] H.M.Lin, X.X.Liu, J.Liu and N.Qu, Communicating and mobile systems: the Pi-calculus, Qinghua Press, Beijing, 2009, pp.28-31.

[9] S.Basu, T.Bultan and m.Ouederni, Deciding choreography realizability, In POPL 2012, 2012, pp.191-202.

[10] O.Grumberg, O.Kupferman and S.Sheinvald, Variable automata over infinite alphabets, In LNCS, Springer,   vol 6031, 2010, pp.561-572.

[11] L.Chen and R.Chow, Building A Valid Web Service Composition Using Integrated Service Substitution and Adaptation, Journal of Information and Decision Sciences, vol.2, no.2, 2010, pp.113-131.

[12] L.Chen, Yan Li and R.Chow, Enhancing Web Service Registries with Semantics and Context Information, Proceedings of 7th Intl. Conference on Services Computing, 2010, pp.641-644.

[13] A.Segev and E.Toch, Context-Based Matching and Ranking of Web Services for Composition, IEEE Transactions on Services Computing, 2(3), 2009, pp.210-222.

[14] A.Pahlevan and H.A.Muller, Self-Adaptive Management of Web Service Discovery, Proceedings of the PhD Symposium at the 8th European Conference on Web Services, 2010, pp.21-24.

[15] A.K.Tripathy and M.R.Patra, Cost Effective, Requirement Oriented Web Service Composition and Adaptation, International Journal on Recent Trends in Engineering and Technology, 2011, pp.95-98.

[16] B.Rochwerger et al, Reservoir-when one cloud is not enough, IEEE Computer 44(3), 2011, pp.44-51.

[17] X.Li, Y.Fan, Q.Z.Sheng, Z.Maamar, H.Zhu, A petri net approach to analyzing behavioral compatibility and similarity of web servicesw, IEEE Transactions on Systems, 2011, pp.1-12.

[18] S.Bourne, C.Szabo, Q.Z.Sheng, Ensuring well-formed conversations between control and operational behaviors of web services, In Proc of the 10th International Conference on Service-Oriented Computing, Shanghai, China, 2012.

[19] V.Sharma, M.Kumar, Web Service Discovery Research: A Study of Existing Approaches,
10